\documentclass[%
 reprint,
 amsmath,amssymb,
 aps,
]{revtex4-2}

\usepackage{graphicx}% Include figure files
\usepackage{dcolumn}% Align table columns on decimal point
\usepackage{bm}% bold math
\usepackage{color}
\begin{document}

\preprint{APS/123-QED}

\title{ The transport properties of Kekulé-ordered graphene $p$-$n$ junctions}% Force line breaks with \\

\author{Peipei Zhang\textsuperscript{1}}

\author{Chao Wang\textsuperscript{2}}
\author{Yu-Xian Li\textsuperscript{1}}

\author{Lixue Zhai\textsuperscript{1}}
\email{zhailixue@126.com}

\author{Juntao Song\textsuperscript{1}}
\email{jtsong@hebtu.edu.cn}
\affiliation{%
$^{1}$College of Physics and Hebei Advanced Thin Films Laboratory, Hebei Normal University, Shijiazhuang, Hebei 050024, China\\
$^{2}$College of Physics, Shijiazhuang University, Shijiazhuang, Hebei 050035, China}%

\date{\today}% It is always \today, today,

\begin{abstract}
The transport properties of electrons in graphene $p$-$n$ junction with uniform Kekulé lattice distortion have been studied using the tight-binding model and the Landauer-Büttiker formalism combined with the nonequilibrium Green's function method. In the Kekulé-ordered graphene, the original $K$ and $K^{\prime}$ valleys of the pristine graphene are folded together due to the $\sqrt{3} \times \sqrt{3}$ enlargement of the primitive cell. When the valley coupling breaks the chiral symmetry, special transport properties of Dirac electrons exist in the Kekulé lattice. In the O-shaped Kekulé graphene $p$-$n$ junction, Klein tunneling is suppressed, and only resonance tunneling occurs. In the Y-shaped Kekulé graphene $p$-$n$ junction, the transport of electrons is dominated by Klein tunneling. When the on-site energy modification is introduced into the Y-shaped Kekulé structure, both Klein tunneling and resonance tunneling occur, and the electron tunneling is enhanced. In the presence of a strong magnetic field, the conductance of O-shaped and on-site energy-modified Y-shaped Kekulé graphene $p$-$n$ junctions is non-zero due to the occurrence of resonance tunneling. It is also found that the disorder can enhance conductance, with conductance plateaus forming in the appropriate range of disorder strength. The ideal plateau value is found only in the Kekulé-Y system. 
\end{abstract}

\maketitle

%\tableofcontents
\section{\label{sec:level1}INTRODUCTION}

Since the successful fabrication of graphene in 2004, the research of graphene has been paid more and more attention \cite{ref1,ref2,ref3,ref4,ref5,ref6,ref7,ref8,ref9,ref10,ref11}. This is mainly due to graphene's unique energy band structure, which has a linear dispersion relation $(E=\pm \hbar v|\textbf{k}|)$ at low energy, allowing the properties of the electrons to be described by the massless relativistic Dirac equation \cite{ref12}. The unique energy band structure of graphene leads to many peculiar properties, such as the anomalous quantum Hall effect \cite{ref9}, Klein-Gordon tunneling \cite{ref10}, and the Hall plateau assumes the half-integer values for the $p$-$p$ or $n$-$n$ junction, new plateaus emerge at $e^{2} / h$ and $(3 / 2) e^{2} / h$ for the $p$-$n$ junction under the disorder \cite{ref2,ref3,ref7}. As is known, the conduction and valence bands of graphene touch each other at the corners of the hexagonal Brillouin zone, which is divided into two inequivalent groups, labeled as the $K$ or $K^{\prime}$ valley. Therefore, in addition to the usual charge and spin degrees of freedom, electrons in graphene also have additional valley degrees of freedom \cite{ref13,ref14}. This leads to a new discipline rising as valleytronics \cite{ref13,ref14,ref15,ref16}.

The low-energy excitations of graphene with the opposite chirality at valleys $K$ and $K^{\prime}$. Chirality $\hat h$ is a fundamental property for relativistic massless Dirac fermions, which is defined by the projection of spin $\pmb{\sigma}$ onto momentum $\pmb{\hat p}$, namely, $\hat h = \frac{1}{2}{\pmb{\sigma}\cdot {\pmb{\hat p} \mathord{\left/{\vphantom {{\hat p} {\left| {\hat p} \right|}}} \right. \kern-\nulldelimiterspace} {\left| \pmb{\hat p} \right|}}}$. Chiral symmetry breaking, namely, coupling of Dirac fermions with opposite chiralities, leads to dynamical mass generation for elementary particles. Recently, chiral symmetry breaking has been successfully demonstrated experimentally in graphene with Kekulé lattice distortion \cite{ref17}, where the lattice distortion version of graphene with periodicity \cite{ref18,ref19}. The unit cell in the Kekulé (Kek) graphene superlattice is enlarged by $\sqrt{3} \times \sqrt{3}$, so that the original $K$ and $K^{\prime}$ valleys of the pristine graphene are folded into the $\Gamma$ point, the wave vector $\mathbf{G}=\mathbf{K}-\mathbf{K}^{\prime}$ \cite{ref18,ref19,ref20,ref21,ref22,ref23,ref24}. There are two kinds of bond distortions in Kek graphene, namely Y-shape and O-shape, called Kek-Y and Kek-O textures. The Y-shaped periodic alternation of weak and strong bonds is called Kek-Y ordering, and the O-shaped periodic alternation is called Kek-O ordering. The Kek-O and Kek-Y superlattices have the same Brillouin zone. With the successful experimental fabrication of the Kek structures \cite{ref17,ref25,ref26}, a series of intriguing phenomena in this structure were discovered, such as topological effects in the electron and phonon spectra  \cite{ref23,ref27,ref28}, valley supercurrent \cite{ref29}, Klein tunneling \cite{ref30}, valley precession \cite{ref31}, enhanced Andreev reflection \cite{ref32}, and so on. In addition, the coupling term of the valley is complex, which affects the characteristics of the valley. Therefore, studying the transport properties of valley-coupled graphene $p$-$n$ junctions is necessary. It helps to understand the effect of valley coupling on transport properties and phenomena related to the valley.

In this paper, using the tight-binding model and the Landauer-Büttiker formalism combined with the nonequilibrium Green's function method \cite{ref4,ref7,ref33,ref34,ref35,ref36}, we investigate the electron transport for both clean and disordered samples at the $p$-$n$ junction of Kek graphene. The Kek graphene has two structures, Kek-O and Kek-Y. For the Kek-O structure, the energy band opens a gap due to chiral symmetry breaking. For the Kek-Y structure, the chiral symmetry is unbroken, and the energy band is gapless. However, when the on-site energy modification term exists, the characteristics of one valley are destroyed and a single-valley phase is formed. The numerical results show that resonance tunneling is occurring in the Kek-O $p$-$n$ junction. Nevertheless, the electron tunneling is mainly the Klein tunneling in the Kek-Y $p$-$n$ junction, which has a similar procedure to the original graphene $p$-$n$ junction. While the on-site energy modification term is introduced into the Kek-Y structure, resonance tunneling and Klein tunneling will coexist, and electron tunneling will be enhanced correspondingly. 

The behavior of Kek graphene $p$-$n$ junction conductance in the magnetic field could be classified into four regimes: quantum tunneling, transition regime, quasi-classical regime, and quantum Hall regime. In the presence of a strong magnetic field, the conductance of O-shaped and on-site energy-modified Y-shaped Kekulé graphene $p$-$n$ junctions is non-zero due to the occurrence of resonance tunneling. After adding disorder into the $p$-$n$ junction, the value of conductance in the Kek-O structure will reach a plateau at suitable disorders, but it is lower than the ideal value. In the Kek-Y structure new plateaus [with values ${e^2/h, (3/2)e^2/h}$] can appear in a wide range of disorder strength $W$ \cite{ref7}. In the Kek-Y structure with the on-site energy modification $U \ne 0$, the lowest plateaus [with values ${e^2/h}$] emerge for $W=2$. For large filling factors, higher-order plateaus can not be formed. At large disorder, it will drive the systems into the insulating regime.

The remainder of this paper is organized as follows. In Sec. \ref{SC2}, we introduce the Hamiltonian that describes the Kek structure in a lattice model and the formulas for calculating electron transport. In Sec. \ref{SC3}, we present and discuss our main results. A brief conclusion is presented in the last section.

\section{MODEL AND FORMULA}\label{SC2}

We consider a graphene sheet with a uniform Kek deformation \cite{ref17,ref26,ref37} on the $x$-$y$ plane, in which the two graphene valleys are coupled. The Kek-O structure composed of alternative bond modulations like a benzene-ring molecule is taken into account as schematically shown in Fig. 1(a) and is obtained by intercalating Li to monolayer graphene on SiC substrate \cite{ref17}. The Kek-Y structure has a Y-shaped periodic alternation of weak and strong bonds as shown in Fig. 1(b), where one of the six Carbon atoms in a supercell to sit on the substrate-atom (Cu) vacancy that makes three neighbor C-C bonds shorten \cite{ref26}. Here, we use the most general tight-binding representation \cite{ref18,ref37,ref38}, the Hamiltonian of the Kek graphene $p$-$n$ junction [see Figs. 1(a) and 1(b)] are given by:
\begin{eqnarray}
\begin{split}
\label{Eq1}
H &= \sum\limits_{im} {\left( {{\varepsilon _{im}} + {U_{i2}}} \right)} c_{im}^\dag {c_{im}} - {t_0}\sum\limits_{\langle im,jn\rangle } {{e^{i{\phi _{im,jn}}}}} c_{im}^\dag {c_{jn}} \\&- {t_1}\sum\limits_{\langle {i}m,{j}n\rangle } {{e^{i{\phi _{{i}m,{j}n}}}}} c_{{i}m}^\dag {c_{{j}n}} ,
\end{split}
\end{eqnarray}
where the first term accounts for the on-site energy, $i(j)$ is a supercell index, $m(n)$ is the atomic index, $m(n)$ = $1$-$6$ corresponds to atoms ${A_1}$, ${A_2}$, ${A_3}$, ${B_1}$, ${B_2}$, ${B_3}$ respectively. $\sum\limits$ runs over all adatom sites and $ c_{im}^\dag ({c_{im}})$  is the creation (annihilation) operator at the site $im$. In the left and right leads, ${\varepsilon _{im}}={E_L}$ or ${E_R}$, which can be controlled by the gate voltages. The potential drop from the right to the left leads is assumed to be linear, i.e., ${\varepsilon _{im}} = k({E_R} - {E_L})/(6M + 1) + {E_L} + {w_{im}}$, where $M$ is the length of the center region and $k = 1,2,3...6M$ [see Figs. 1(a) and 1(b)]. With the change of ${E_L}$ and ${E_R}$, the central area presents an $n$-$n$ (${E_L}$, ${E_R}<0$) or $p$-$n$ (${E_L}<0,{E_R}>0$) state. The on-site disorder energy ${w_{im}}$ is uniformly distributed in the range $\left[ { - {W \mathord{\left/{\vphantom {W 2}} \right.\kern-\nulldelimiterspace} 2},{W \mathord{\left/{\vphantom {W 2}} \right.\kern-\nulldelimiterspace} 2}} \right]$ with the disorder strength $W$. The disorder exists only in the center region. $U$ represents the on-site energy modification of the carbon atoms associated with the substrate Cu vacancies, which is present only at $m=2$ in the Kek-Y structure. The second and third terms in Hamiltonian are nearest-neighbor hopping. Due to the modification of the hopping energy in the band Hamiltonian, it contains ${\delta_0={-\delta}t}$ and ${\delta_1=2{\delta}t}$. Here represents the corresponding energy modification to the electron's hopping as shown in Figs. 1(a) and 1(b), ${t_0} = (1 -{\delta})t$, ${t_1} = (1 + 2\delta )t$. In Kek-O, ${\delta}$ affects the size of the band gap. In Kek-Y, ${\delta}$ affects the slope of the two cones \cite{ref34}, and the on-site energy modification affects the size of the band gap in the single-valley phase \cite{ref29}. When an external magnetic field $(0, 0, B)$ is applied perpendicular to the Kek graphene sheet, a phase ${\phi _{ij}}$ is added in the hopping term, and ${\phi _{ij}} = \int_i^j {\vec A \cdot d\vec l/{\phi _0}}$ with the vector potential $\vec A = ( - By,0,0)$ and ${\phi _0} = \hbar /e$. 

\begin{figure}[htbp]
\begin{center}
\includegraphics[width=0.48\textwidth]{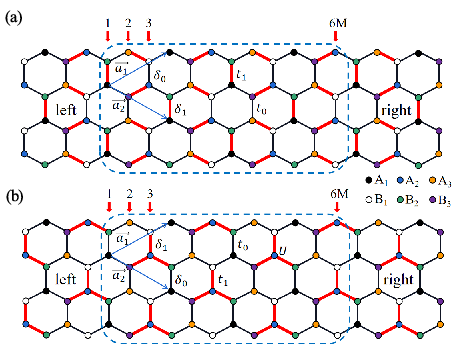}
\end{center}
\caption{\label{fig1} (Color online) The schematic diagram for zigzag edge Kek graphene $p$-$n$ junction. (a)Kek-O graphene, (b)Kek-Y graphene. ${\delta_\alpha }$ $(\alpha=0,1)$ represents the corresponding energy modification to the electron’s hopping, and $U$ is for possible site-energy modification. $\vec{a_1}$ and $\vec{a_2}$ are the basis vectors of the supercell. The size of the center region (blue dots line area) is described by the width $N$ and length $M$. The parameters $N=2, M=2$.}
\end{figure}

The energy band structure of Kek graphene nanoribbon with a zigzag edge was obtained from Eq. (1). The energy band structure of the three systems is shown in Fig. 2, respectively. In Figs. 2(a)-(c), we can be seen clearly that the Kek-O order opens a gap in the Dirac cone through coupling valleys \cite{ref18,ref19,ref21,ref25,ref28}. However, the Kek-Y order does not impose a mass on the Dirac fermions and still has chiral symmetry \cite{ref17,ref38,ref39,ref40}. Once on-site energy at a special position on the Kek-Y structure occurs, the chiral symmetry for one of the valleys is broken, the appearance of the single Dirac cone together with a flat band across the Dirac point. The Dirac points must appear in pairs for the two-dimensional electron systems described by the Weyl equation. The single Dirac cone on the Kek-Y lattice with $U \ne 0$ stems from the fact that the present Dirac fermions do not satisfy the two-component Weyl equation due to intervalley coupling and inversion symmetry-breaking \cite{ref38}. Fig. 2(c) shows an energy gap near $E=0$, caused by the size effect. Under the vertical magnetic field, they form different Landau energy levels, shown in Figs. 2(d)-(f). Kek-O does not have the zeroth Landau energy level, and Kek-Y has the zeroth Landau energy level. When $U=0.5t$, Landau energy levels split, the upper flat band shows a little down-warping dispersion, so within the bending depth, the longitudinal conductance is nonzero, whereas the Hall conductivity is zero \cite{ref41}.

\begin{figure}[htbp]
\begin{center}
\includegraphics[width=0.48\textwidth]{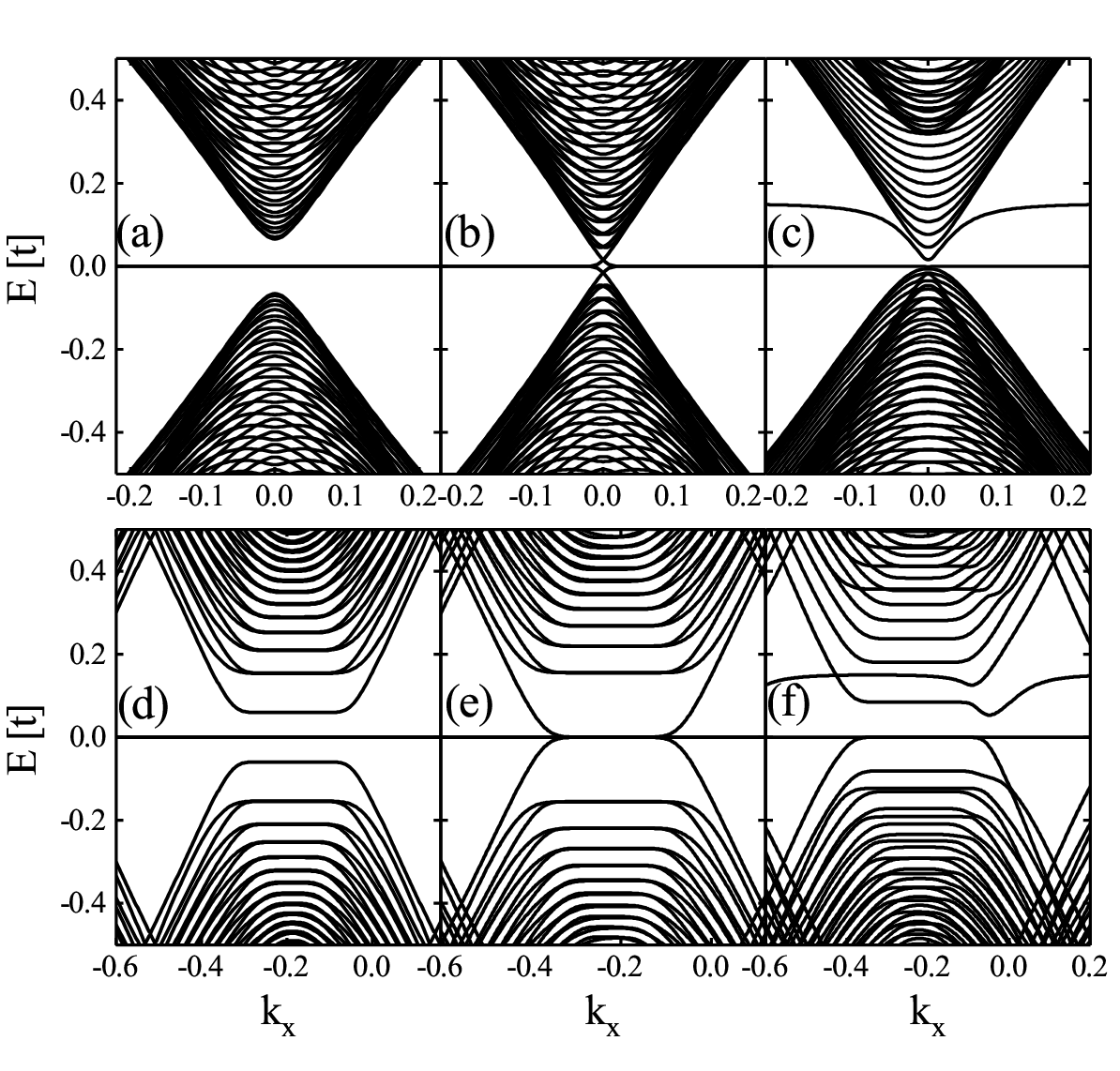}
\end{center}
\caption{\label{fig2} The energy band structure of the Kek graphene. (a)(d) Kek-O graphene, (b)(e) Kek-Y graphene with $U=0$, (c)(f) Kek-Y graphene with $U=0.5t$. In the upper panel are bands with no applied magnetic field, while in the lower panel are bands consider a perpendicular magnetic field with ${\phi}=0.007$. The ribbon width $N=50$.}
\end{figure}

With this Kek $p$-$n$ junction system, the current can be obtained from the Landauer-Büttiker formula \cite{ref42}: 
\begin{eqnarray}\label{Eq2}
J = \left( {2e/h} \right)\int {d\varepsilon {T_{LR}}(\varepsilon )} \left[ {{f_L}(\varepsilon ) - {f_R}(\varepsilon )} \right],
\end{eqnarray}
where ${f_\alpha }(\varepsilon ) = 1/\{ \exp [(\varepsilon  - e{V_\alpha })/{k_B}T] + 1\} (\alpha  = L, R)$ is the Fermi distribution function in the left and right Kek graphene leads. Here, ${T_{LR}}\left( \varepsilon  \right) =$ Tr$\left[ {{\Gamma _L}{G^r}{\Gamma _R}{G^a}} \right]$ is the transmission coefficient with the linewidth functions ${\Gamma _\alpha }\left( \varepsilon  \right) = i\left[ {\Sigma _\alpha ^r(\varepsilon ) - \Sigma _\alpha ^a(\varepsilon )} \right]$. The retarded Green's function is ${G^r}\left( \varepsilon  \right) = {\left[ {{G^a}\left( \varepsilon  \right)} \right]^\dag } = {\left[ {\varepsilon I - H - \Sigma _L^r - \Sigma _R^r} \right]^{ - 1}}$, where $H$ is the Hamiltonian matrix of the central region and $I$ is the unit matrix. The retarded self-energy function $\Sigma _\alpha ^r(\varepsilon )$ coupling to lead $\alpha$ can be obtained from $\Sigma _\alpha ^r(\varepsilon ) = {H_{c,\alpha }}g_\alpha ^r\left( \varepsilon  \right){H_{\alpha ,c}}$, where ${H_{c,\alpha }}$ is the coupling from the central region to lead $\alpha $, and similarly ${H_{\alpha ,c}}$ couples from lead $\alpha $ to the central region. $g_\alpha ^r\left( \varepsilon  \right)$ is the surface retarded Green’s function for semi-infinite lead $\alpha $ calculated using the transfer-matrix method \cite{ref43}. After obtaining the current $J$, the linear conductance is given by $G = {\lim _{V \to 0}}{{dJ} \mathord{\left/{\vphantom {{dJ} {dV}}} \right.\kern-\nulldelimiterspace} {dV}}$.

In the following numerical calculations, we use the hopping energy $t = 2.75$ eV as the energy unit. The corresponding energy modification to the electron’s hopping $\delta  = 0.02t$. The electron transport process only occurs near the Dirac point, where Klein tunneling is most likely to occur, so we take the value of Fermi energy $E_\alpha$ as far less than the nearest transition energy in the original graphene structure, that is, ${E_\alpha } \ll t$. The width $N$ is chosen as $N =50$ in all calculations. Since the nearest-neighbor carbon-carbon distance is $a = 0.142$ nm, the width is $\left( {3N - 1} \right)a \approx 21.2$ nm for $N=50$. The magnetic field is expressed in terms of $\phi $ with $\phi  \equiv {{\left( {{{3\sqrt 3 } \mathord{\left/
 {\vphantom {{3\sqrt 3 } 4}} \right.\kern-\nulldelimiterspace} 4}} \right){a^2}B} \mathord{\left/{\vphantom {{\left( {{{3\sqrt 3 } \mathord{\left/{\vphantom {{3\sqrt 3 } 4}} \right. \kern-\nulldelimiterspace} 4}} \right){a^2}B} {{\phi _0}}}} \right.\kern-\nulldelimiterspace} {{\phi _0}}}$  is the magnetic flux in the honeycomb lattice and the flux quanta ${\phi _0} = \hbar /e$. The on-site energy modification $U=0.5t$. In the case of disorder, the conductance is averaged over up to 1200 random. 
\section{NUMERICAL RESULTS AND DISCUSSION}\label{SC3}

We first study the clean Kek graphene junction. Fig. 3 shows the conductance $G$ versus the Fermi level of right lead ${E_R}$ for ${E_L}=-0.1t$ and ${E_L}=-0.2t$. When ${E_R} < 0$, the center region is an n-n junction. $G$ is approximatively quantized and exhibits a series of plateaus. The value of the plateaus depends on the number of the transverse sub-bands of the lead with finite widths, which can be seen in Figs. \ref{fig2}(a)-(c). For ${E_R} < {E_L}$, due to the fixed sub-band numbers in the left region, no higher plateaus appear. However, the on-site energy modification term is special when it exists.  As shown in Fig. 3(e), when ${E_R}=-0.15t$, the conductance decreases because the number of the fixed subbands in the right region decreases. When ${E_R}>0$, the center region is the $p$-$n$ junction and the electron transport from the conduction band to the valence band, so it is a tunneling process. In previous works, it has been certified that the transporting process of the electron in the graphene $p$-$n$ junction is Klein tunneling \cite{ref7}.

\begin{figure}[htbp]
\begin{center}
\includegraphics[width=0.5\textwidth]{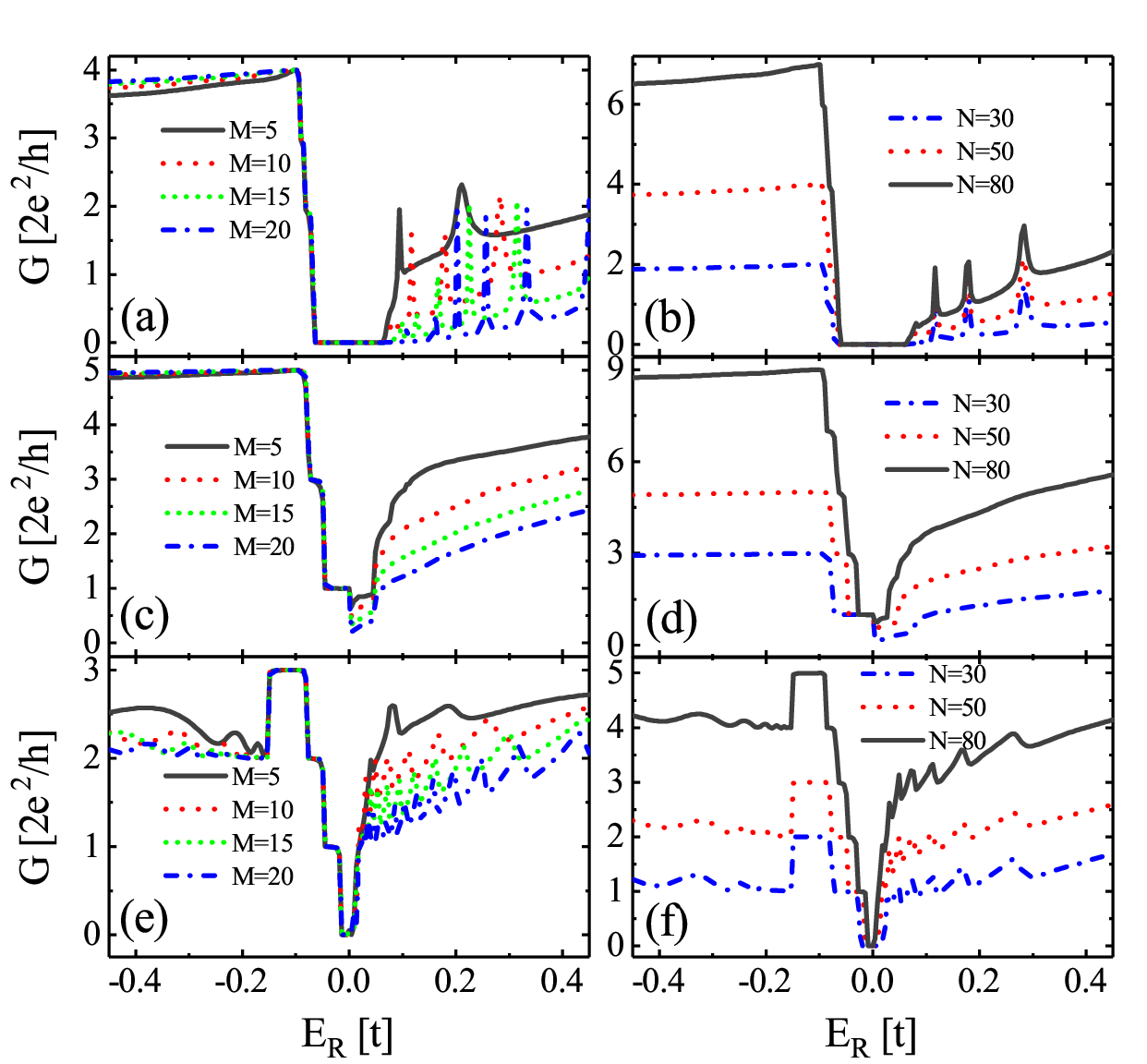}
\end{center}
\caption{\label{fig3} (Color online) The conductance $G$ vs ${E_R}$ for the different Kek graphene at $W=0$, ${\phi}=0$, ${E_L}=-0.1t$. (a)(b) Kek-O graphene, (c)(d) Kek-Y graphene with $U=0$, (e)(f) Kek-Y graphene with $U=0.5t$. The conductance in the left panel is obtained at center region width $N=50$ with different center region lengths, whereas the conductance in the right panel is obtained at center region length $M=10$ with different center region widths.}
\end{figure}

In Figs. \ref{fig3}(a) and 3(b), we can see that in the Kek-O $p$-$n$ junction, the value of conductance oscillates clearly with ${E_R}$ increasing, and a series of peaks appear. When the width of the central region is fixed, resonance tunneling resonance is more likely to occur as the length of the central region increases, thus increasing the number of resonance peaks. The position of resonance peaks varies with the length of the central region, as shown in Fig. 3(a). When the length of the central region is fixed, the number of subbands in the left region increases as the width of the central region increases, which increases the values of resonance peaks, and the position of resonance peaks remains unchanged, as shown in Fig. 3(b). The number and position of resonance peaks are closely related to the length of the junction. This is the feature of resonance tunneling. This is different from Klein tunneling in graphene $p$-$n$ junction due to the chiral symmetry breaking in the energy band of the Kek-O structure, as shown in Fig. \ref{fig2}(a). Furthermore, near $E_R=0$, the conductance $G=0$, which is due to the opening of the band gap in the energy band structure of Kek-O. 

In Figs. \ref{fig3}(c) and 3(d), the conductance $G$ with changing $E_R$ in the Kek-Y struction is presented. Different from the Kek-O structure, there is less oscillating behavior in the curve of conductance. With $E_R$ increasing from $0$ to $0.45t$, the conductance rises but is always less than the corresponding plateau value in the $n$-$n$ region. With the increase of $M$, the probability of scattering increases, and the conductance are slightly weakened, as shown in Fig. 3(c). With the increase of $N$, the number of subbands in the left region increases, leading to an increase in conductance, as shown in Fig. 3(d). From these, it is certified that electron transport in the $p$-$n$ junction of Kek-Y graphene is mainly Klein tunneling \cite{ref7}. Due to the chiral symmetry being unbroken by the coupling of the two valleys in the Kek-Y structure, a gapless linear energy band remains, as shown in Fig. \ref{fig2}(b). When valley coupling does not change the characteristics of the two valleys, the transport properties of the Kek $p$-$n$ junction are similar to those of graphene. 

Then the on-site energy modification of the Kek-Y structure is adjusted to be $U=0.5t$, the energy band of which is shown in Fig. \ref{fig2}(c). The gapless linear bands and the gapped parabolic bands exist together. Only one valley breaks the chiral symmetry, and the other one keeps the chiral symmetry invariant. The single Dirac cone here is quite different from that in graphene, even though they have linear dispersion relation \cite{ref44}. In Figs. \ref{fig3}(e) and 3(f), the conductance $G$ with changing $E_R$ in the Kek-Y struction of $U=0.5t$ is presented. We can see that when $E_R$ changes from $0$ to $0.05t$, compared with the Figs. \ref{fig3}(c) and 3(d), the conductance raise more quickly. Continuing to increase $E_R$ to $0.45t$,  we find that the curves of conductance present upward behaviors slowly but oscillate strongly. With the increase of $M$, the number and position of resonance peaks vary. With the increase of $N$, the resonance peaks' values increase while the peak's position remains unchanged. In other words, the resonance tunneling and the Klein tunneling occur at the same time when the electron is transported into the Kek-Y $p$-$n$ junction with $U=0.5t$. This is because the chiral symmetry is broken in only one of the two valleys. Owing to the single Dirac cone of massless Dirac fermions, the transport of electrons on the Kek-Y lattice with $U \ne 0$ has a new feature.

\begin{figure}[htbp]
\begin{center}
\includegraphics[width=0.5\textwidth]{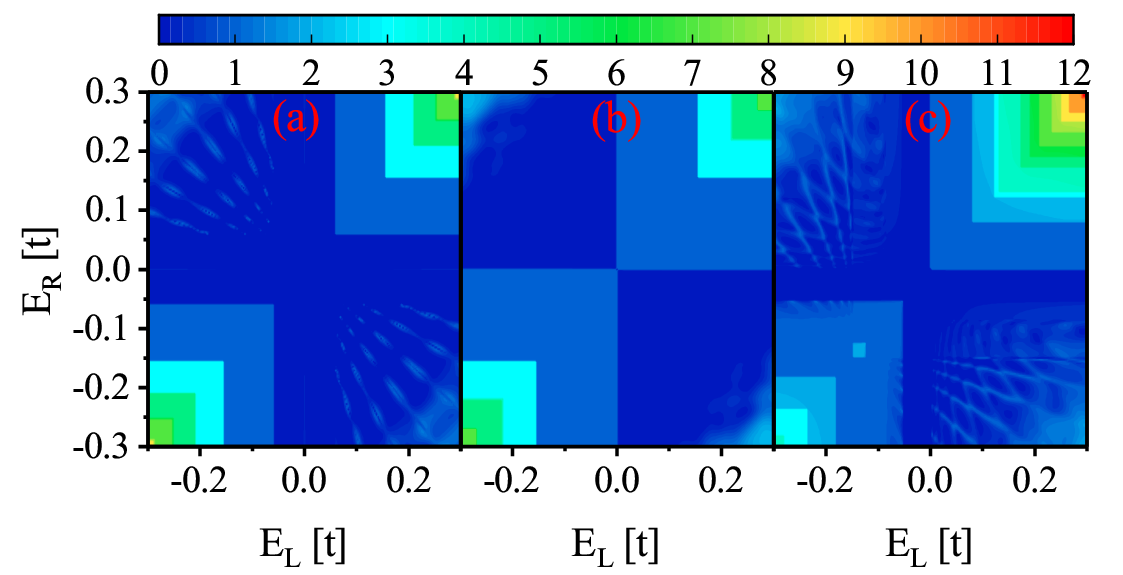}
\end{center}
\caption{\label{fig4} (Color online) The conductance $G$ (in the unit of $2{{{e^2}} \mathord{\left/{\vphantom {{{e^2}} h}} \right. \kern-\nulldelimiterspace} h}$) vs ${E_L}$ and ${E_R}$ with $M=10$, ${\phi}=0.007$, $W=0$. (a) Kek-O graphene, (b) Kek-Y graphene with $U=0$, (c) Kek-Y graphene with $U=0.5t$.}
\end{figure}

Next, with a vertical magnetic field being introduced, the electron should travel along the edge of the device in a clockwise or anti-clockwise direction. The conductance of the clean samples is calculated shown in Fig. 4. Under the large magnetic field strengths, the sub-bands evolve into the Landau levels. The conductance plateaus in the $n$-$n$ region evolve into the Hall plateau. The plateau values are given by $\min \left( {\left| {{\nu _L}} \right|,\left| {{\nu _R}} \right|} \right){{{e^2}} \mathord{\left/{\vphantom {{{e^2}} h}} \right. \kern-\nulldelimiterspace} h}$, where ${\nu _\alpha }$ is the filling factors in the lead $\alpha $ \cite{ref45,ref46}. For example, in the Kek-O $n$-$n$ region, when ${E_L} =  - 0.1t$ $({\nu _L} =  - 2)$ and ${E_R} = - 0.2t$ $({\nu _R} = - 6)$, the conductance value at $2e^2/h$ is shown in Fig. 4(a).

In the $p$-$n$ region (${E_L} < 0$ and ${E_R} > 0$), instead of the conductance plateaus, the conductance shows the periodic oscillations with the change of $E_\alpha$ [see Fig. 4]. This is because the Hall edge states of electrons and holes are formed in the $p$ and the $n$ regions under strong magnetic fields. The transport direction of electrons is opposite to that of holes. In the $p$-$n$ junction, the presence of resonant tunneling and the formation of mixed states at the interface between the Hall edge states of electrons and holes in different regions leads to conductance. In addition, the magnetic field breaks the chiral symmetry, but there is a corresponding dressed chiral symmetry. In the $p$-$n$ junction of Kek-Y with $U=0$, it is more difficult to form resonance tunneling, and only mixed states exist, therefore the conductance is smaller.

\begin{figure}[htbp]
\begin{center}
\includegraphics[width=0.5\textwidth]{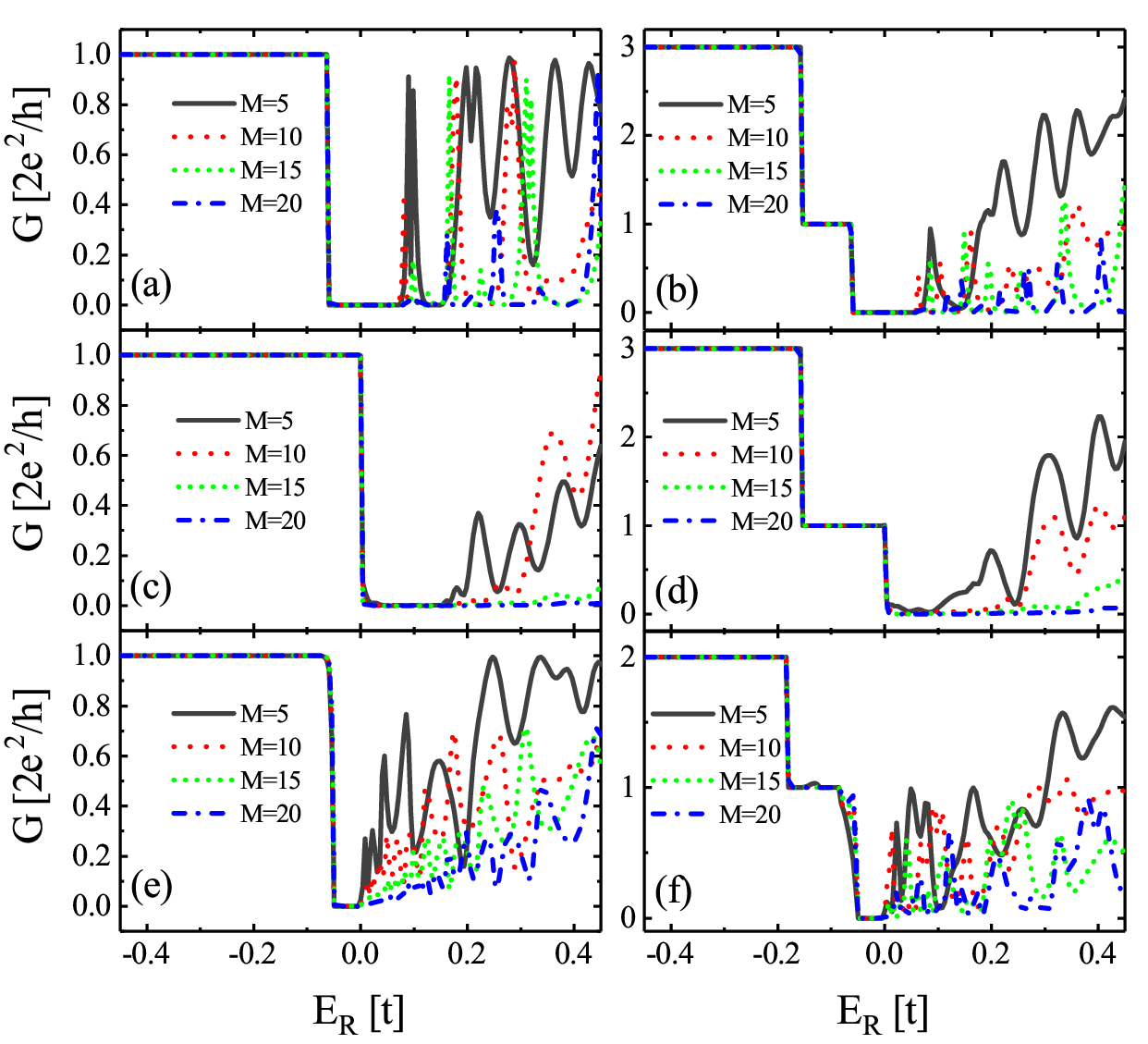}
\end{center}
\caption{\label{fig5} (Color online) When ${\phi}=0.007$, the conductance $G$ vs ${E_R}$ of graphene with different central lengths and difference Kek structure. In the left panels ${E_L}=-0.1t$, in the left panels ${E_L}=-0.2t$. (a)(b) Kek-O graphene, (c)(d) Kek-Y graphene with $U=0$, (e)(f) Kek-Y graphene with $U=0.5t$.} 
\end{figure}

The curves of conductance with $E_R$ changing at different $M$ are shown in Fig. \ref{fig5}. To illustrate how the conductance change under the magnetic field in different structures, $E_L$ and ${\phi}$ are set to $-0.1t$, $-0.2t$, and $0.007$, respectively. In the $n$-$n$ region, the values of the conductance are given by $\min \left( {\left| {{\nu _L}} \right|,\left| {{\nu _R}} \right|} \right){{{e^2}} \mathord{\left/{\vphantom {{{e^2}} h}} \right. \kern-\nulldelimiterspace} h}$, which stems from the lowest Landau levels of the left or right terminal. When $E_R>0$, the center region of the devices is changed to be the $p$-$n$ junction under the magnetic field, so the curves show oscillating behaviors clearly. The irregular oscillations are mainly due to the contribution of $G$ by the resonant tunneling and the mixed Hall edge states. In (a)-(d) of Figs. 5, we can see that the conductance is zero near $E_R=0.05t$, but the causes of zero conductance are different. In Figs. 5(a) and (b), the $G=0$ is caused by the band gap. In Figs. 5(c) and (d), when the filling factor is small, the carriers are more likely to form localized states and mixed states are difficult to form, resulting in a smaller conductance. As the filling factor increases, the conductance increases. 

Moreover, the oscillation strength of the conductance $G$ depends on the length of the $p$-$n$ junction $M$. As the increase of $M$, the oscillation strength decay, and the conductance $G$ decreases. This is because the Hall edge states for electrons and holes are well separated in space and cannot form mixture states, which leads to a reduction in conductance. In particular, for the Kek-Y structure with $U=0$, there is no resonance tunneling occurs due to the presence of dressed chiral symmetry, and the conductance $G$ is strongly suppressed at large junction length $M=20$, and $G$ is close to zero [see Figs. 5(c) and 5(d)]. However, the conductance in the Kek-O and Kek-Y with $U=0.5t$ systems is not zero for the same junction length $M$ [see Figs. 5(a), 5(b), 5(e), and 5(f)]. This shows that in the $p$-$n$ junction with Kek-O and Kek-Y ($U \ne 0$) structures, the mixed states of electrons and holes Hall edge states disappears with increasing $M$ and the conductance is contributed by resonant tunneling. This is consistent with the results in Fig. 4. 

\begin{figure}[htbp]
\begin{center}
\includegraphics[width=0.5\textwidth]{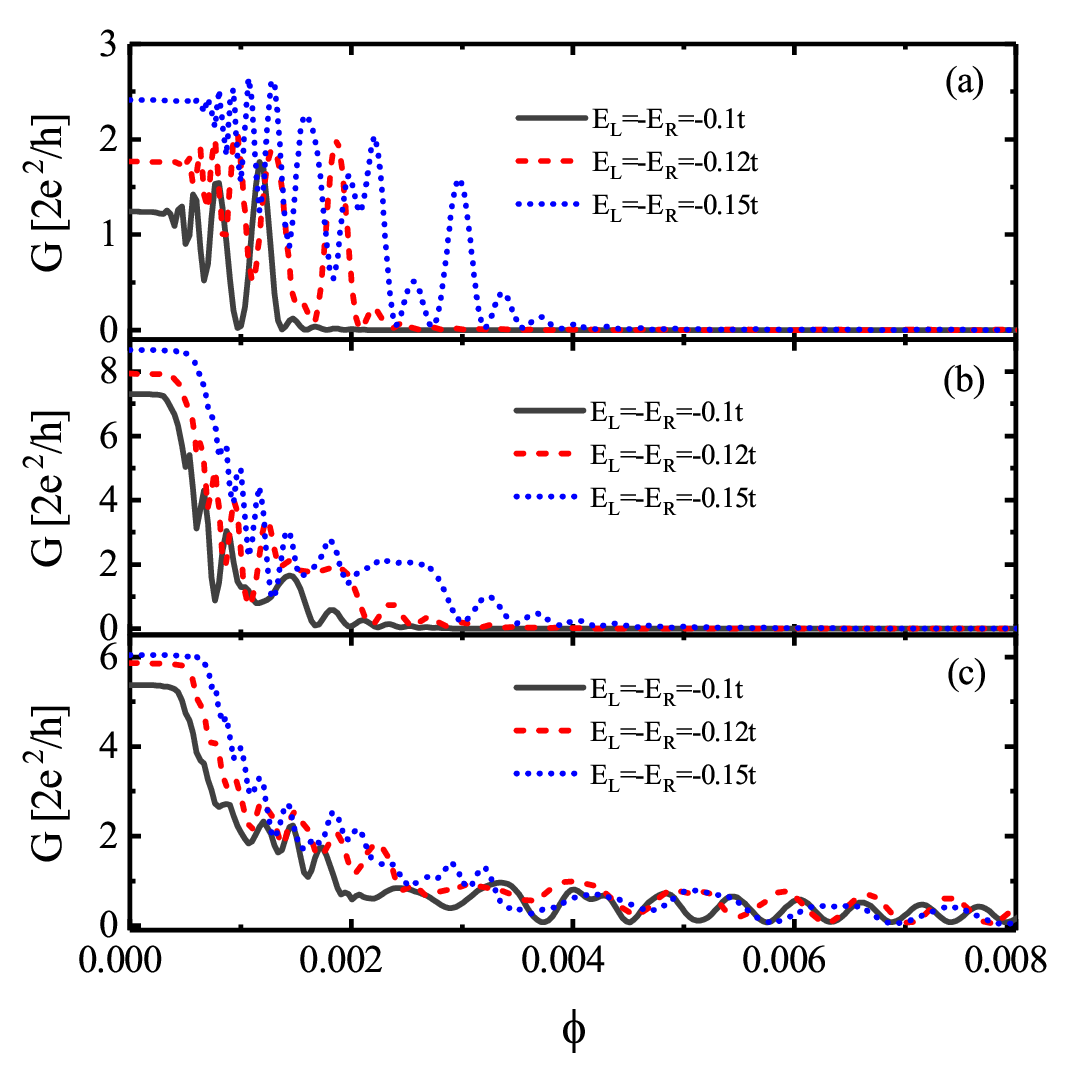}
\end{center}
\caption{\label{fig6} (Color online) The conductance $G$ vs ${\phi}$ of Kek graphene with different $V_0$. (a) Kek-O graphene, (b)Kek-Y graphene with $U=0$, (c) Kek-Y graphene with $U=0.5t$. The other parameters $N=150$, $M=10$.}
\end{figure}

In the following discussion, we learned that the behavior of Kek graphene $p$-$n$ junction conductance in the magnetic field could be classified into four regimes: quantum tunneling, transition regime, quasi-classical regime, and quantum Hall regime \cite{ref45}. It is investigated how these four mechanisms are determined by the relevant parameters. If we adopt $E_L$ = $-E_R$, we use $V_0$ = $E_L$ = $-E_R$ for simplicity. In Kek structures, the variation curves of conductance with magnetic field strength at different $V_0$ are shown in Fig. 6. 

The variation curve of conductance with magnetic field in the Kek-O $p$-$n$ junction is given in Fig. 6(a). At a weak magnetic field, due to the large radius of motion of the incident electron, the boundary states are unable to form. The transmission is contributed by quantum tunneling of Landau levels. The conductance in the $p$-$n$ junction has some series of oscillatory changes. First, the Kek-O lattice structure forms a small number of Landau levels under a weak magnetic field, and as the magnetic field increases, the Landau levels do not cross the Fermi surface, but the degeneracy is enhanced, leading to an increased tendency for oscillations in conductance. Subsequently, the $G$ decreases and shows a Shubnikov-de Haas oscillation. As ${\phi}$ increases, the Landau levels shift upward through the Fermi level, resulting in a decrease in $G$. When the Landau levels at both terminals are around the Fermi level, the resonant peak appears.

When the magnetic field strength increases, the coexistence of resonance tunneling, snake states and Aharanov–Bohm (AB) effect, the oscillation of $G$ shows an irregular pattern. When the magnetic field strength increases more, the snake states are formed in the $p$-$n$ interface. $G$ versus ${\phi}$ shows an oscillation and both the period and the magnitude of the oscillation decrease as ${\phi}$ is enhanced. When ${\phi}$ becomes very large, the Hall edge states of electrons and holes are formed in the $p$ and the $n$ regions, respectively. However, the transport direction of electrons is opposite to that of holes, so the conductance is very small. In addition, the oscillatory behavior of $G$-${\phi}$ is influenced by the potential variation in the central region. As $V_0$ increases, all curve changes move to larger values of ${\phi}$. 

The oscillation of conductance in Kek-Y is similar to that in Kek-O $p$-$n$ junction. The difference is that the Kek-Y structure forms at weak magnetic fields with a high number of Landau levels, and instead of enhancing the $G$ as the magnetic field increases, it decreases and exhibits a Shubnikov-de Haas oscillation. By comparing Fig. 6(b) and 6(c), we find that there is a clear periodic oscillation of the conductance as ${\phi}$ in the presence of a strong magnetic field at $U \ne 0$. Here the oscillations are caused by warp dispersion in the energy band structure. The conductance oscillation decays with the increasing magnetic field, but it has a higher electron tunneling probability than that with $U=0$. 

Now we study the effect of disorders on conductance. In Fig. 7, we can see that the conductance versus ${E_R}$ at fixed ${E_L=-0.1} t$ $({\nu _L}=-2)$ for the different disorders. The dashed lines of different colors represent the ideal value ($\left[ {{{\left| {{\nu _L}} \right|\left| {{\nu _R}} \right|} \mathord{\left/ {\vphantom {{\left| {{\nu _L}} \right|\left| {{\nu _R}} \right|} {\left( {\left| {{\nu _L}} \right| + \left| {{\nu _R}} \right|} \right)}}} \right.\kern-\nulldelimiterspace} {\left( {\left| {{\nu _L}} \right| + \left| {{\nu _R}} \right|} \right)}}} \right]{{{e^2}} \mathord{\left/ {\vphantom {{{e^2}} h}} \right. \kern-\nulldelimiterspace} h}$) of conductance in the $p-n$ region under different filling factors \cite{ref46,ref47}. In the $n$-$n$ region, $G$ is less influenced by the disorders for $W \leq 1$, keeping the Hall plateaus at the value of $\min \left( {\left| {{\nu _L}} \right|,\left| {{\nu _R}} \right|} \right){{{e^2}} \mathord{\left/{\vphantom {{{e^2}} h}} \right. \kern-\nulldelimiterspace} h}$. If the disorder strength $W$ is increased to $2$ or more, the conductance $G$ decrease clearly, reaching to be zeros with $W=6$. In the $p$-$n$ regions, considering the effect of the disorders, that is the mix of the electron and hole edge states, the conductance $G$ can be enhanced in a suitable range of disorder strength. At large disorder, $G$ is very small for all $E_L$ and $E_R$, because it is much more difficult to completely mix all the states, the system enters the insulating regime before the occurrence of the complete state mixing. However, there are some differences between the conductance enhancements in the $p$-$n$ region of different structures.   

\begin{figure}[htbp]
\begin{center}
\includegraphics[width=0.5\textwidth]{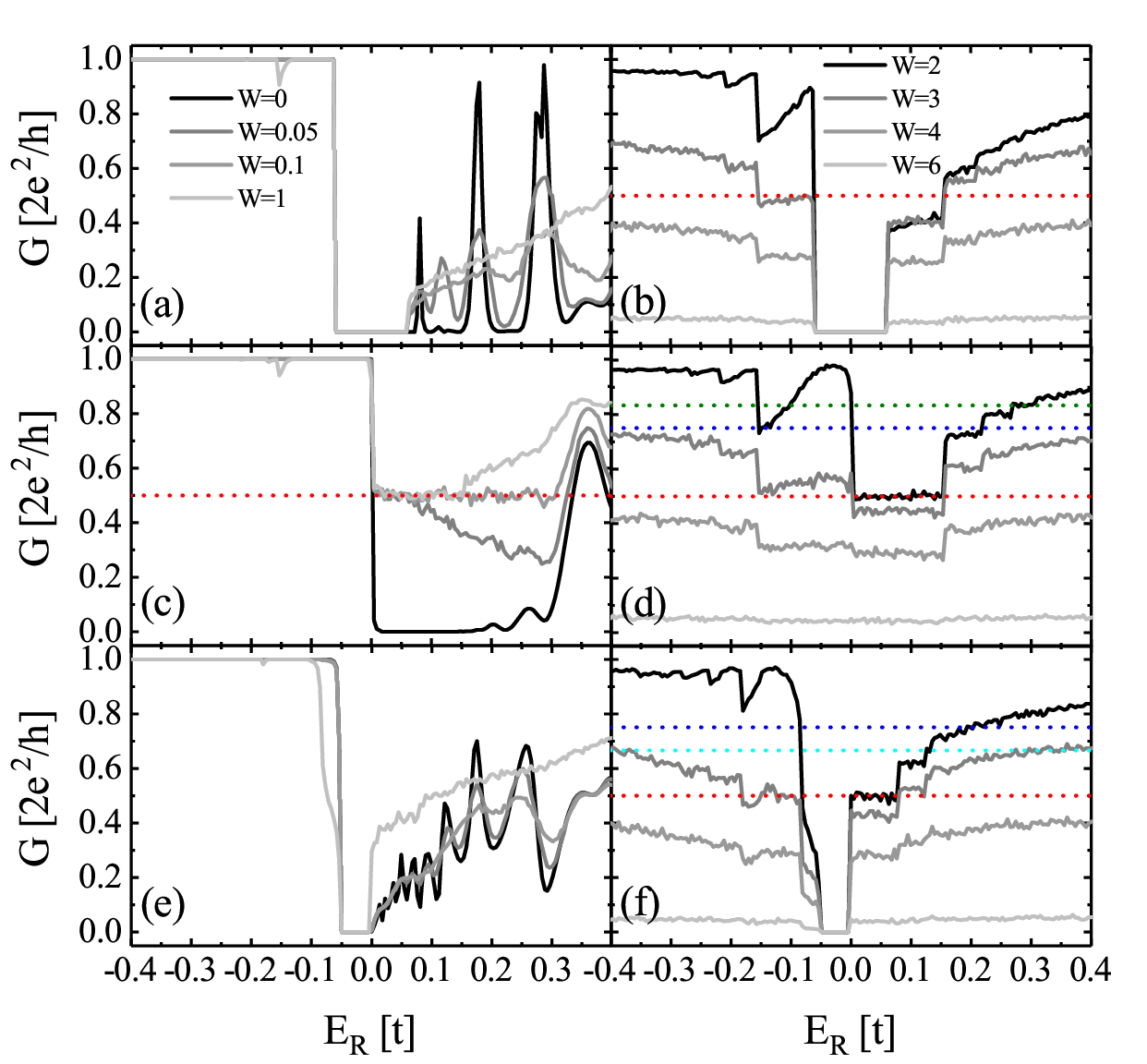}
\end{center}
\caption{\label{fig7} (Color online) The conductance $G$ vs ${E_R}$ for the different disorder strengths and different Kek graphene at ${\phi}=0.007$. (a)(b) Kek-O graphene, (c)(d) Kek-Y graphene with $U=0$, (e)(f) Kek-Y graphene with $U=0.5t$. The conductance in the left panel is obtained of disorder strength $W$ from 0 to 1. While in the right panels, the conductance is obtained from the disorder strength $W$ of 2 to 6. The parameters ${E_L}=-0.1t$, $M=10$.}
\end{figure}

In Figs. 7(a) and 7(b), for the Kek-O $p$-$n$ junction, the sharp peaks of $G$ slowly becomes flat with $W$ increasing from $0$ to $1$. In particular, when the disorder intensity is in the range of 2 to 4, the conductance plateaus are formed around the special value of ${E_R}$, where the value corresponds to the conductance peak with $W=0$. But the value of the plateaus is lower than the ideal value. For example, for ${E_L} = - 0.1t$ (${\nu _L} = -2$) and ${E_R} = 0.1t$ (${\nu _R} = 2$), the ideal conductance reaches the plateau of ${e^2/h}$ showed as the red dotted line in Fig. 7(b), but the actual value is $0.8{e^2/h}$ for $W=2$ (or 3). For the Kek-O $p$-$n$ junction, the conductance of disorder-induced enhancement is difficult reaching the ideal value.

Figs. 7(c) and 7(f) depict the conductance versus $E_R$ at different disorder strengths for the Kek-Y structure with $U=0$. With the increase of $W$ form 0, the conductance $G$ in the $p$-$n$ regions is strongly enhanced even for very small $W$. The lowest conductance plateau with ${\nu_L}=-2$ and ${\nu_R}=2$ is well established with its plateau value at ${e^2/h}$ for $W=0.1$. This plateau remains for a broad range of disorder strength $W$ (from 0.1 to 2). For higher filling factors, the conductance is also enhanced by the disorder. But it is difficult to reach the ideal plateaus value, and even unable to form plateaus. For example, for ${\nu_L} = -2$ and ${\nu_R} = 6$, the ideal conductance reaches the plateau of ${1.5e^2/h}$ showed as the blue dotted line in Fig. 7(d), that is slightly higher than the actual value when $W=2$. 

For the Kek-Y structure with $U=0.5t$, the conductance versus $E_R$ at different disorder strengths which is shown in Figs. 7(e) and 7(f). The lowest conductance plateau value is still at ${e^2/h}$ for $W=2$, but the plateau remains for a small range of disorder intensity. The conductance is difficult to reach the ideal plateau value for higher filling factors. When  ${\phi}=0.007$, $W \ne 0$, the transport properties of the Kek-Y $p$-$n$ junction are similar to those of graphene, but for the on-site energy modification $U=0.5t$, the effect of disorder-induced transport enhancement is weakened. 

\section{CONCLUSIONS}\label{SC4}

In summary, using the nonequilibrium Green's function method combined with the tight-binding Hamiltonian, we study the transport properties in graphene $p$-$n$ junctions with the Kek distortion, under the clean and disordered systems. It is found that in the Kek-O graphene $p$-$n$ junction, Klein tunneling is suppressed and only resonance tunneling occurs due to the broken chiral symmetry. In the Kek-Y $p$-$n$ junction, the transport of the electrons is dominated by Klein tunneling due to pseudospin conservation. In addition, the characteristics of a valley will be destroyed once the on-site energy modification term is introduced into the Kek-Y structure. Because of the presence of the single-valley phase, resonance tunneling and Klein tunneling will coexist, and electron tunneling will be enhanced correspondingly. Under strong magnetic fields, the conductance of Kek-Y ($U=0$) graphene $p$-$n$ junctions is approximately zero due to the presence of dressed chiral symmetry without resonant tunneling occurring. In a word, the conductance of electron transport in the Kek-Y system with a single Dirac cone is higher than those in the system with the Dirac cones appearing in pairs. On the other hand, in the $p$-$n$ region, the conductance is enhanced at a suitable disorder and suppressed at a large disorder. Our work provides new ideas for the study of heterojunction structures. 

\begin{acknowledgments}
This work was supported by the National Natural Science
Foundation of China (Grant No.\ 11874139), the Natural
Science Foundation of Hebei province (Grant No.\ A2019205190).
\end{acknowledgments}

\end{document}